\documentclass[epj,nopacs]{svjour}
\usepackage{graphics}
\usepackage{epsfig}
\def\lsim{\raise0.3ex\hbox{$<$\kern-0.75em\raise-1.1ex\hbox{$\sim$}}}
\def\gsim{\raise0.3ex\hbox{$>$\kern-0.75em\raise-1.1ex\hbox{$\sim$}}}

\def\mean#1{\left<#1\right>}
\def\Journal#1#2#3#4{{#1}{\bf #2} (#4) #3}
\def\IJMPA{{Int. J. Mod. Phys. A}}
\def\EPJC{{Eur. Phys. J. C}}
\def\JPG{{J. Phys. G}}

\def\NPA{{Nucl. Phys. A}}
\def\NPB{{Nucl. Phys. B}}
\def\PLB{{Phys. Lett. B}}
\def\PLC{Phys. Repts.\ }

\def\PRL{Phys. Rev. Lett.\ }
\def\PRD{{Phys. Rev. D}}
\def\PRC{{Phys. Rev. C}}
\def\PR{Phys. Rev.\ }

\def\ZPC{{Z. Phys. C}}
\def\ARNPS{{Ann. Rev. Nucl. Part. Sci.\ }} 
\def\RMP{Rev. Mod. Phys.\ }

\def\RPP{Rep. Prog. Phys.\ }
\begin{document}
\title{Successes and failures with hard probes}
\author{M.~J.~Tannenbaum 
\thanks{Research supported by U.S. Department of Energy, DE-AC02-98CH10886.}
}                     
%
%
\institute{Brookhaven National Laboratory\\Upton, NY 11973-5000 USA}
\date{Received: date / Revised version: date}
%
\abstract{   The two major pillars of searches for the Quark Gluon Plasma have been: J/$\Psi$ suppression, proposed in 1986, and 
 observed at both SPS fixed target energies and at RHIC; and, more recently, the suppression of $\pi^0$ with $p_T\geq 3$ GeV/c by a factor $\sim 5$ in Au+Au central collisions, observed at RHIC in 2001, which had been predicted in advance as a consequence of  Landau-Pomeranchuck-Migdal coherent (gluon) bremsstrahlung by the outgoing hard-scattered partons traversing the medium. However, new effects were discovered and the quality of the measurements greatly improved so that the clarity of the original explanations has become obscured. For instance: J/$\Psi$ suppression is the same at SpS and RHIC. Is it the QGP, comovers, something else? QCD provides beautiful explanations of $\pi^0$ and direct $\gamma$ measurements in p-p collisions but precision fits of the best theories of $\pi^0$ suppression barely agree with the Au+Au data. Better data are needed for $10< p_T <20$ GeV/c, systematic errors are needed in theory calculations, the values of parameters of the medium such as $\mean{\hat{q}}$ derived from precision fits are the subject of controversy. Baryons are much less suppressed than mesons, leading to an anomalous $\bar{p}/\pi$ ratio for $2\leq p_T\leq 4.5$ GeV/c, but beautiful theoretical explanations of the effect such as recombination do not work in detail. Heavy quarks seem to be suppressed the same as the light quarks, naively arguing against the bremsstrahlung explanation and suggesting exotic, possible transformational explanations.  Di-hadron correlations reveal a trigger side ridge, possible Mach cones on the away side, vanishing and reappearance of away jets, both wide and normal jet correlations with and without apparent loss of energy. Can this all be explained consistently?  Preliminary results of direct $\gamma$ production in Au+Au appear to indicate a suppression approaching that of $\pi^0$ for $p_T\approx 20$ GeV/c and a possibly thermal component for $1\leq p_T\leq$ 3 GeV/c. What are the implications? Are fragmentation photons a problem? Regeneration of direct $\gamma$ by outgoing partons is predicted, leading to negative $v_2$---is there evidence for or against it? STAR and PHENIX have different observations relevant to the existence of monojets in d+Au collisions. Will new data clarify the situation? When? etc. These and other issues will be discussed with a view to identify which conclusions are firm and where further progress towards real understanding is required.   
}
%
%
\authorrunning{M.~J.~Tannenbaum}
\titlerunning{Successes and Failures with Hard Probes}
\maketitle
%
\section{Introduction}
\label{sec:intro}
     The motivation to study nuclear matter under extreme conditions of temperature and density reached in Relativistic Heavy Ion conditions is primarily the search for a new state of matter, called the Quark Gluon Plasma for historical reasons because it was thought to be a gas of deconfined quarks and gluons covering a volume much larger than an individual nucleon (a plasma being an ionized gas)~\cite{MJTROP}. The early signals, searched for at the Berkeley Bevalac, the Brookhaven AGS and the CERN SpS, were collective hydrodynamic flow~\cite{ROPBeV,CP99}, strangeness enhancement~\cite{RM82,E802-90}, baryon stopping~\cite{Ahle98} and $J/\Psi$ suppression~\cite{MatsuiSatz,NA50} which was believed to be the `gold-plated' signature of deconfinement. All these signatures were found at the nucleon-nucleon c.m. energy range $2<\sqrt{s_{NN}}< 17.2$ GeV of these machines but in my opinion the Quark Gluon Plasma was not found~\cite{MJTROP}. 
Nevertheless, the prevailing opinion in the field until the end of the 20$^{\rm th}$ century was that the `Hard Probe', $J/\Psi$ suppression, was the best method to find the QGP at RHIC. 
     
\begin{figure}[!h]
\includegraphics[width=0.45\textwidth]{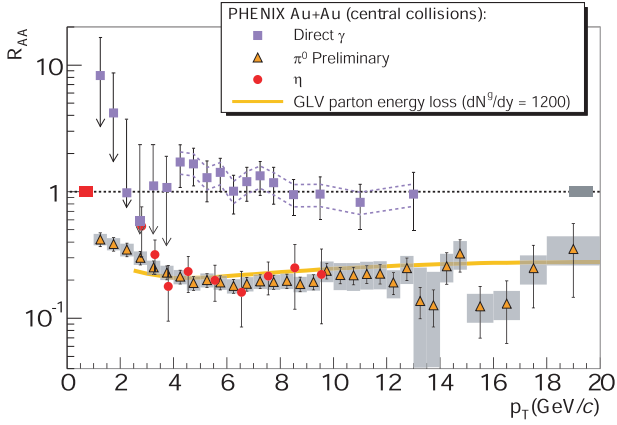} 
\caption[]{Nuclear modification factor, $R_{AA}$ for direct-$\gamma$, $\pi^0$ and $\eta$ in Au+Au central collisions at $\sqrt{s_{NN}}=200$ GeV~\cite{YAQM05}, together with GLV theory curve~\cite{GLV}.}
\label{fig:QM05wow}
\end{figure}

\begin{figure}[!h]
\includegraphics[width=0.90\linewidth]{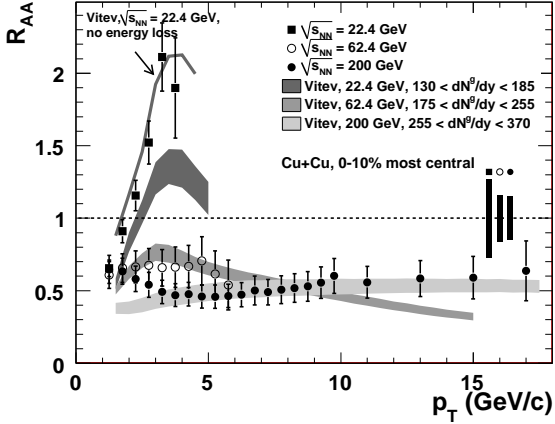} 
\caption[]{Nuclear modification factor, $R_{AA}$ for $\pi^0$ in Cu+Cu central collisions at $\sqrt{s_{NN}}=200$, 62.4 and 22.4 GeV~\cite{ppg084}, together with Vitev theory curves~\cite{Vitev2}.}
\label{fig:PXCu}
\end{figure}

     However, in the 1990's a new hard probe of the color response of the medium, `Jet Quenching'~\cite{GP90,WG92} was proposed and given a firm basis in QCD~\cite{BDMPS} as coherent Landau-Pomeranchuck-Migdal coherent bremsstrahlung of gluons by the outgoing hard-scattered partons traversing the medium. The discovery of a huge quenching of high $p_T$ $\pi^0$ by a factor of ~5 in central Au+Au collisions at RHIC (Fig.~\ref{fig:QM05wow})
coupled with the absence of such an effect at SpS energies (Fig.~\ref{fig:PXCu})  
has led to the opinion by some that we now really understand everything that is happening in RHI collisions at RHIC (e.g. nearly opaque matter, with partons visible only from the surface). In my opinion this 
is the principal failure at RHIC. In fact, it is clear to me that we are still on a long learning curve and far from understanding in detail the many discoveries and effects observed at RHIC, the underlying fundamental physics of QCD in a color-charged medium, and the properties of the medium produced at RHIC. I will sketch a few of the successes and outline some of the many open questions below. 

\section{Successes}
\begin{figure}[ht] 
\begin{center}
\begin{tabular}{c}
\hspace*{-0.0\linewidth}\includegraphics[width=0.70\linewidth]{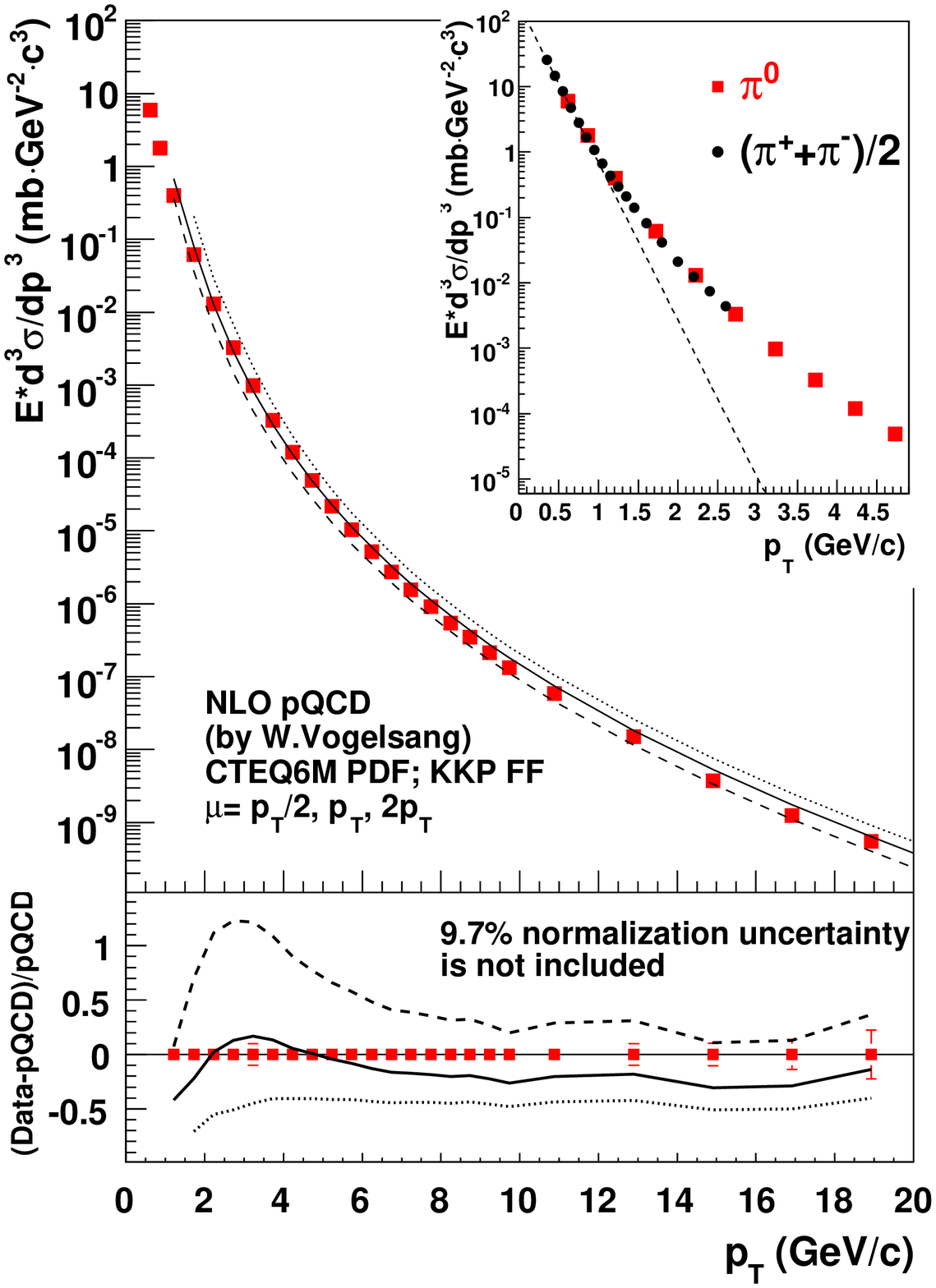}\cr
\hspace*{-0.10\linewidth}\includegraphics[width=0.70\linewidth,height=0.60\linewidth]{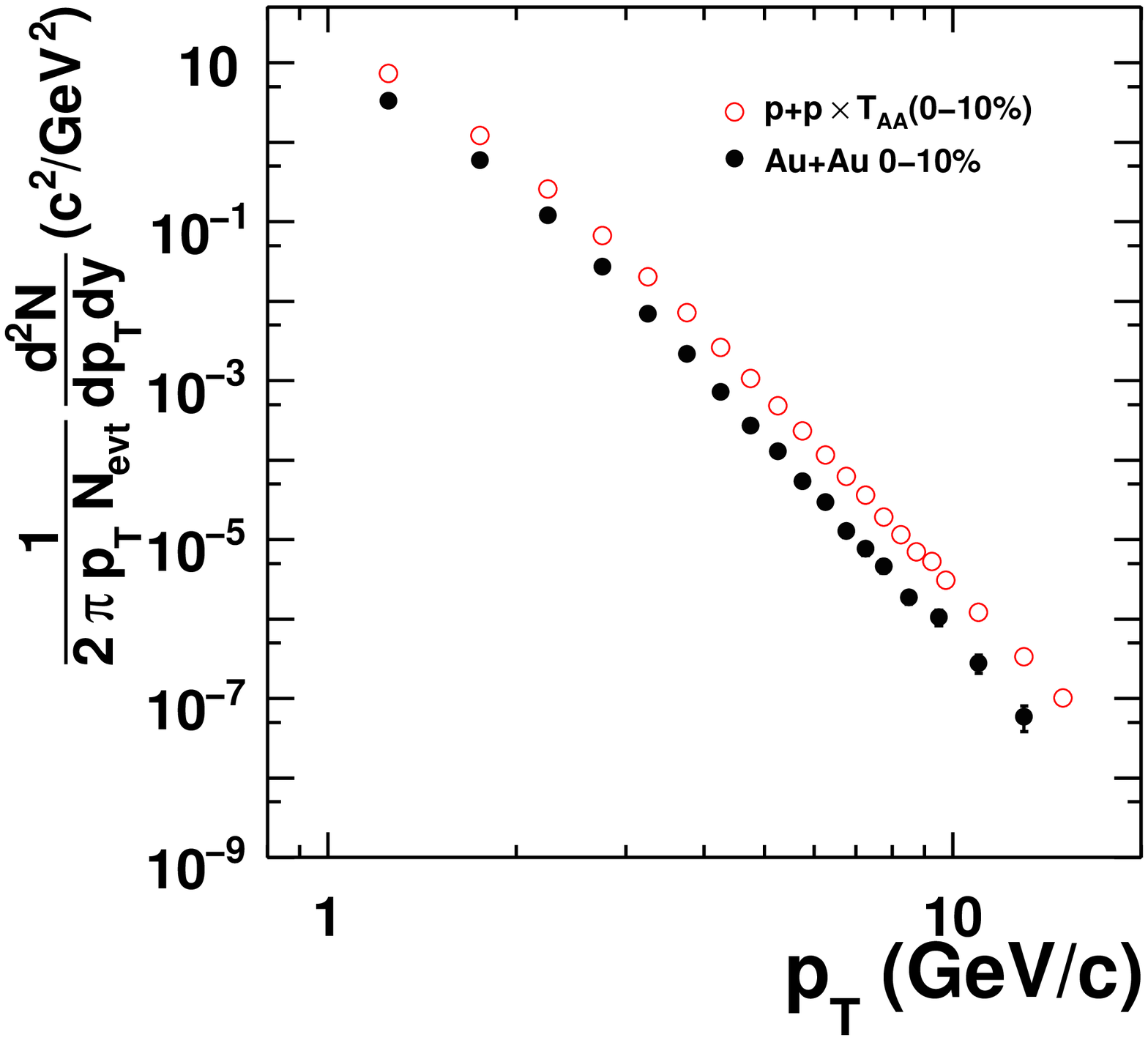} 
\end{tabular}
\end{center}
\caption[]
{a) (top) Invariant cross section at mid-rapidity for $\pi^0$ production in p-p collisions at $\sqrt{s}=200$ GeV~\cite{PXPRD76}. b) (bottom) log-log plot of semi-inclusive invariant yield of $\pi^0$ in central (0-10\%) Au+Au collisions at $\sqrt{s_{NN}}=200$ GeV~\cite{ppg054} and Invariant cross section for p-p collisions~\cite{PXppPRL91} multipied by $\mean {T_{AA} (0-10\%)}$    
\label{fig:pi0cross} }
\end{figure}

     The real success at RHIC is the precision and accuracy of the measurements and excellent data sets at the same $\sqrt{s_{NN}}$, with absolute cross sections and semi-inclusive yield measurements in p-p, Au+Au, (see Fig.~\ref{fig:pi0cross}) d+Au and Cu+Cu, as well as measurements over a broad range of $\sqrt{s_{NN}}$ for Cu+Cu and p-p. The impressive agreement of the p-p measurements with QCD {\em predictions} (Fig.~\ref{fig:pi0cross}a) gives added confidence to both the measurements and the theory. 
     
     In Fig.~\ref{fig:pi0cross}b, both the p-p and Au+Au spectra exhibit a pure power law for $p_T>4$ GeV/c with $n=8.10\pm 0.05$ which indicates that their ratio will be constant $\sim 0.2$ over the range $4 \leq p_T\leq 12$ GeV/c.       
     The ratio:  
     \begin{equation}
     R_{AA}(p_T)=\frac{d^2 N^{\pi}_{AA}/dp_T dy N^{inel}_{AA}}{ \mean{T_{AA}} d^2\sigma^{\pi}_{pp}/dp_T dy}
     \label{eq:RAA}
     \end{equation}
     is used as the most convenient way to represent the physics and not because, for instance, the efficiency cancels in the ratio, which it certainly doesn't because the mean background multiplicity to the processes of interest increases by a factor of $\gsim 300$ from p-p to central Au+Au collisions. 
            
\subsection{Parton suppression at RHIC} 
   Fig.~\ref{fig:QM05wow} shows that at $\sqrt{s_{NN}}=200$ GeV, direct-$\gamma$ which do not interact with the medium are not suppressed while the $\pi^0$ and $\eta$ mesons which are fragments of hard-scattered light-quarks and gluons are suppressed. This indicates a strong medium effect on partons, consistent with QCD LPM energy loss as indicated by the agreement with the theory~\cite{GLV}. I actually think that the data are more consistent with a constant $R_{AA}\sim 0.2$ from $4\leq p_T\leq 20$ GeV/c (as would be given by a constant-fractional energy loss and a pure power-law partonic $p_T$ spectrum~\cite{ppg054}) than with a slowly rising value of $R_{AA}$ with increasing $p_T$ as indicated by the theory and in fact this is borne out by the best fit to the data~\cite{ppg079}. Another new PHENIX result nicely illustrates that parton suppression begins somewhere between $\sqrt{s_{NN}}$=22.4 and 62.4 GeV (Fig.~\ref{fig:PXCu})~\cite{ppg084}, but does not completely rule out parton energy loss at 22.4 GeV although it is very suggestive. 
\subsection{Precision tests of models of parton suppression} 
\begin{figure}[!h]
\begin{center}
\includegraphics[width=0.90\linewidth]{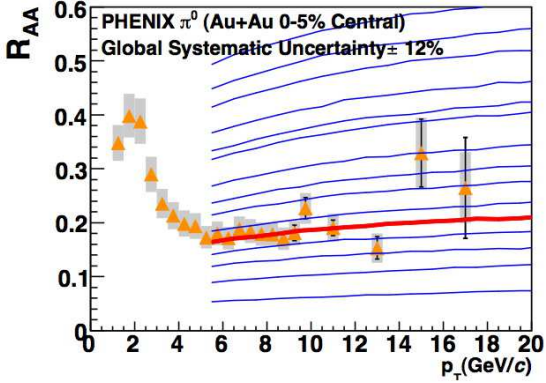} 
\end{center}
\caption[]{PHENIX $\pi^0$ $R_{AA}(p_T)$ for Au+Au central (0-5\%) collisions at $\sqrt{s_{NN}}=200$~\cite{ppg079} compared to PQM model predictions~\cite{PQM}  as a function of $\mean{\hat{q}}$. The thick red line is the best fit. Values of $\mean{\hat{q}}$ corresponding to the lines are shown on Fig.~\ref{fig:RAA20}.  
\label{fig:pi0pqm} }
\end{figure}

\begin{figure}[!h] 
\begin{center}
\begin{tabular}{cc}
\hspace*{-0.02\linewidth}\includegraphics[width=0.51\linewidth]{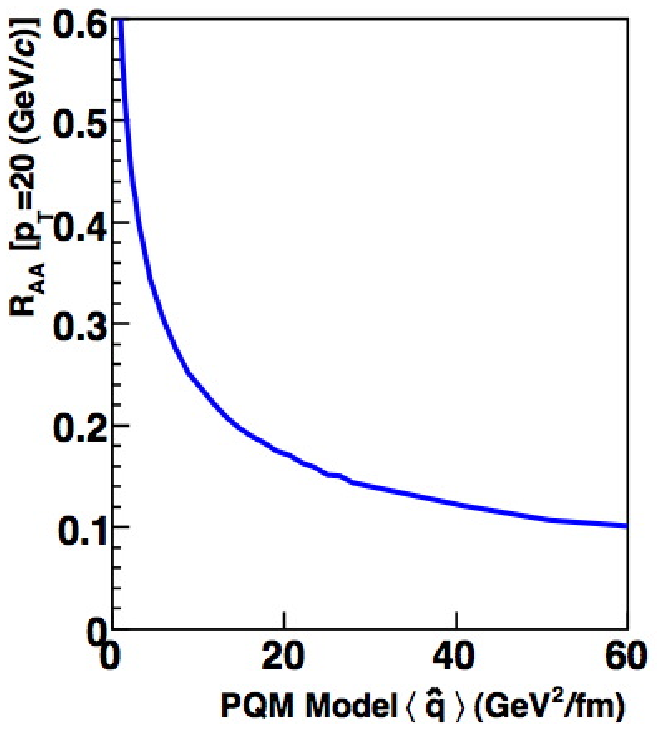} &
\hspace*{-0.08\linewidth}\includegraphics[width=0.53\linewidth]{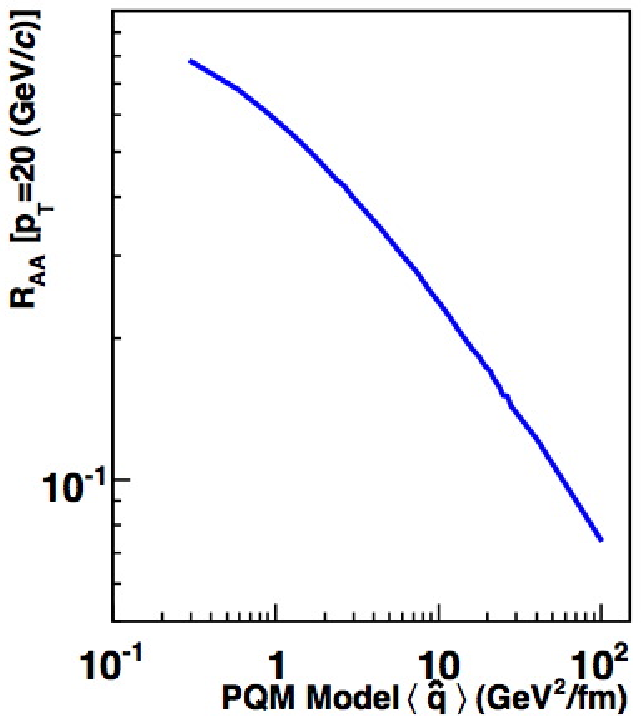}
\end{tabular}
\end{center}
\caption[]
{a) (left) $R_{AA}$ at $p_T=20$ GeV/c as a function of $\mean{\hat{q}}$ in the PQM model~\cite{PQM}. b) (right) same plot on a log-log scale.   
\label{fig:RAA20} }
\end{figure}

   There are many different models of parton suppression with totally different assumptions which all give results in agreement with the PHENIX measurement $R_{AA}^{\pi^0}\approx 0.20$ for $4\leq p_T\leq 20$ GeV/c in Au+Au central collisions. In Jamie Nagle's talk at this meeting, he described how he got all theorists to send him predictions as a function of their main single parameter that characterizes the medium in order to do precision fits to the latest PHENIX  $\pi^0$ data including the correct treatment of correlated experimental systematic errors (Fig.~\ref{fig:pi0pqm} )~\cite{ppg079}. 
 Systematic uncertainties of the theory predictions were not considered. 
 The large value of the transport coefficient $\mean{\hat{q}=\mu^2/\lambda}=13.2^{+2.1}_{- 3.2}$ GeV$^2$/fm from the best fit to the PQM model~\cite{PQM}  (where $\mu$ is the average 4-momentum transfer to the medium per mean free path $\lambda$) is a subject of some debate in both the more fundamental QCD community~\cite{BS06} and the more phenomenological community~\cite{fragility}. For instance it was stated in Ref.~\cite{fragility} that ``the dependence of $R_{AA}$ on $\hat{q}$ becomes weaker as $\hat{q}$ increases'' as is clear from Fig.~\ref{fig:RAA20}a. It was also asserted that ``when the values of the time-averaged transport coefficient $\hat{q}$ exceeds 5 GeV$^2$/fm, $R_{AA}$ gradually loses its sensitivity.'' That statement also appeared reasonable. However, given the opportunity of looking at a whole range of theoretical predictions (kindly provided by the PQM authors~\cite{PQM}) rather than just the one that happens to fit the data, we experimentalists learned something about the theory that was different from what the theorists thought. By simply looking at the PQM predictions on a log-log plot (Fig.~\ref{fig:RAA20}b), it became evident that the PQM prediction could be parameterized as $R_{AA}[p_T=20 {\rm GeV/c}]=0.75/\sqrt{\hat{q}\,({\rm GeV^2/fm})}$ over the range $5<\hat{q}<100$ GeV$^2$/fm. This means that in this range, the fractional sensitivity to $\hat{q}$ is simply proportional to the fractional uncertainty in $R_{AA}$ ($\Delta\hat{q}/\hat{q}=2.0\times \Delta R_{AA}/R_{AA}$), so that improving the precision of $R_{AA}$ e.g. in the range $10\leq p_T\leq 20$ GeV/c will lead to improved precision on $\mean{\hat{q}}$. This should give the theorists some incentive to improve their (generally unstated) systematic uncertainties.  
\subsection{$R_{AA}$ vs. the reaction plane}  
\begin{figure}[!h]
\begin{center}
\includegraphics[width=0.90\linewidth]{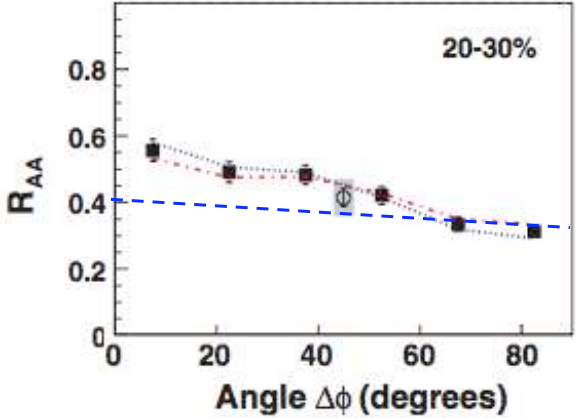} 
\end{center}
\caption[]{$R_{AA}^{\pi^0}$ for $5<p_T< 8$ GeV as a function of $\Delta\phi$ the angle to the reaction plane in Au+Au collisions with centrality 20--30\% at $\sqrt{s_{NN}}=200$ GeV~\cite{ppg054} (data points) compared to prediction for $10<p_T<15$ GeV/c (dashes)~\cite{08053271}. 
\label{fig:th-angle} }
\end{figure}

   Another good synergy between experimentalists and theorists is the study of $R_{AA}$ as a function of angle to the reaction plane and centrality in order to understand the effect of varying the initial conditions (centrality) and the path length through the medium (angle). When PHENIX presented results on $R_{AA}(p_T)$ vs. the angle $\Delta\phi$ to the reaction plane (Fig.~\ref{fig:th-angle})~\cite{ppg054} 
        there was a reaction from the flow community that this is nothing other than a different way to present $v_2$. This is strictly not true for two reasons: 1) $v_2$ measurements are relative while $R_{AA}(\Delta\phi, p_T)$ is an absolute measurement including efficiency, acceptance and all other such corrections; 2) if and only if the angular distribution of high $p_T$ suppression around the reaction plane were simply a second harmonic so that all the harmonics other than $v_2$ vanish (and why should that be?) then $R_{AA}(\Delta\phi, p_T)/R_{AA}(p_T)=1+2 v_2\cos 2\Delta\phi$. In nice talks at this meeting, Steffen Bass and Abhijit Majumder have attempted to put all the theoretical models of jet quenching into a common nuclear geometrical and medium evolution formalism so as to get an idea of the fundamental differences in the models ``evaluated on identical media, initial state and final fragmentation. The only difference in models will be in the Eloss kernel.''~\cite{08053271}. The different models all agreed with the measured $R_{AA}(p_T)$. The agreement with the measured $R_{AA}(\Delta\phi, p_T)$ is not so good (Fig~\ref{fig:th-angle}), but hopefully suggests the way for improvement.  
\subsection{Photons and neutral mesons} 
\subsubsection{Direct photons at low $p_T$}
\begin{figure}[!h]
\begin{center}
\includegraphics[width=0.90\linewidth]{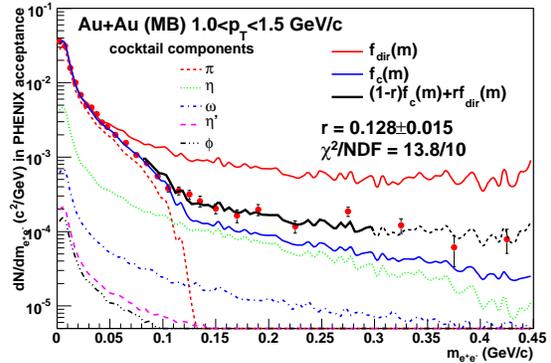} 
\end{center}
\caption[]{Invariant mass ($m_{e+ e^-}$) distribution of $e^+ e^-$ pairs from Au+Au minimum bias events for $1.0< p_T<1.5$ GeV/c~\cite{ppg086}. Dashed lines are Eq.~\ref{eq:KW} for the mesons indicated. Blue solid line is $f_c(m)$, the total di-electron yield from the cocktail of meson Dalitz decays; Red solid line is $f_{dir}(m)$ the internal conversion $m_{e^+ e^-}$ spectrum from a direct-photon ($M>> m_{e^+ e^-}$). Black solid line is a fit of the data to the sum of cocktail plus direct contributions in the range $80< m_{e+ e^-} < 300$ MeV/c$^2$.  
\label{fig:ppg086Fig2KWdist} }
\end{figure}

\begin{figure}[!h]
\begin{center}
\includegraphics[width=0.90\linewidth]{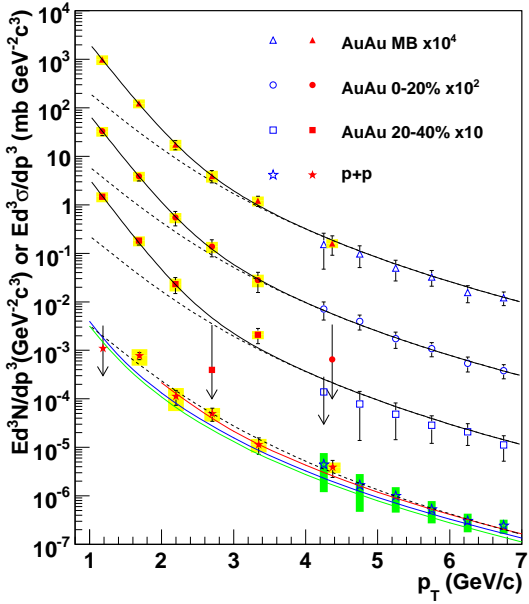} 
\end{center}
\caption[]{Invariant cross section (p-p) or invariant yield (Au+Au) of direct photons as a function of $p_T$~\cite{ppg086}. Filled points are from virtual photons, open points from real photons.
\label{fig:ppg086Fig4} }
\end{figure}

   Internal conversion of a photon from $\pi^0$ and $\eta$ decay is well-known and is called Dalitz decay~\cite{egNPS}. Perhaps less well known in the RHI community is the fact that in any reaction (e.g. $q+g\rightarrow \gamma +q$) in which a real photon can be emitted, a virtual photon (e.g. $e^+ e^-$ pair) of mass $m_{ee}\geq 2m_e$ can be emitted instead. This is called internal-conversion and is generally given by the Kroll-Wada formula~\cite{KW,ppg086}:
   \begin{eqnarray}
   {1\over N_{\gamma}} {{dN_{ee}}\over {dm_{ee}}}&=& \frac{2\alpha}{3\pi}\frac{1}{m_{ee}} (1-\frac{m^2_{ee}}{M^2})^3 \quad \times \cr & &|F(m_{ee}^2)|^2 \sqrt{1-\frac{4m_e^2}{m_{ee}^2}}\, (1+\frac{2m_e^2}{m^2_{ee}})\ ,
   \label{eq:KW}
   \end{eqnarray}
   where $M$ is the mass of the decaying meson or the effective mass of the emitting system. The dominant terms are on the first line of Eq.~\ref{eq:KW}:  the characteristic $1/m_{ee}$ dependence; and the cutoff of the spectrum for $m_{ee}\geq M$ (Fig.~\ref{fig:ppg086Fig2KWdist})~\cite{ppg086}. Since the main background for direct-single-$\gamma$ production is a photon from $\pi^0\rightarrow \gamma +\gamma$, selecting $m_{ee} \gsim 100$ MeV effectively reduces the background by an order of magnitude by eliminating the background from $\pi^0$ Dalitz decay, at the expense of a factor $\sim 1000$ in rate. This allows the direct photon measurements to be extended (for the first time in both p-p and Au+Au collisions) below the value of $p_T\sim 4$ GeV/c, possible with real photons, down to $p_T=1$ GeV/c (Fig.~\ref{fig:ppg086Fig4})~\cite{ppg086}, which is a real achievement. 
The solid lines on the p-p data are QCD calculations which work down to $p_T=2$ GeV/c. The dashed line is a fit of the p-p data to the modified power law $B (1+p_T^2/b)^{-n}$, used in the related Drell-Yan~\cite{Ito81} reaction, which flattens as $p_T\rightarrow 0$. For Au+Au, the exponential spectrum of excess photons above the $\mean{T_{AA}}$ extrapolated p-p fit are suggestive of a thermal source. This is quite distinct from the case for e.g. $\pi^0$ production, where the spectra are exponential in both p-p and Au+Au collisions as $p_T\rightarrow 0$ (Fig.~\ref{fig:pi0cross}). 


\subsubsection{Direct photons and mesons up to $p_T=20$ GeV/c}
\begin{figure}[!h]
\begin{center}
\includegraphics[width=0.90\linewidth]{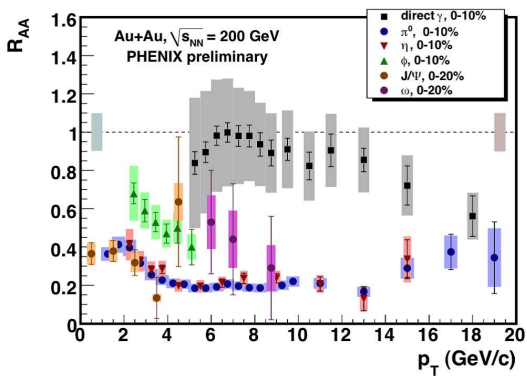} 
\end{center}
\caption[]{$R_{AA}(p_T)$ for direct-$\gamma$ and the mesons indicated in Au+Au central collisions at $\sqrt{s_{NN}}=200$ GeV~\cite{allmQM08}.
\label{fig:raa_mesons_AuAu} }
\end{figure}

   PHENIX continues its relentless pursuit of measuring $R_{AA}$ for pseudo-scalar ($\pi^0$ and $\eta$) and vector ($\omega$, $\phi$, $J/\Psi$) mesons and direct photons over the broadest $p_T$ range (Fig.~\ref{fig:raa_mesons_AuAu})~\cite{allmQM08}. The $\pi^0$ and $\eta$ continue to track each other to the highest $p_T$. The $\phi$ and $\omega$ vector mesons appear to track each other also but with a different value of $R_{AA}(p_T)$. Interestingly, the $J/\Psi$ seems to track the $\pi^0$ for $0\leq p_T\leq 4$ GeV/c; and it will be interesting to see whether this trend continues at higher $p_T$. 
   
   The direct-$\gamma$ case is striking and possibly indicative of trouble ahead for the LHC. With admittedly large systematic errors, which should not be ignored, the direct-$\gamma$ appear to become suppressed for $p_T> 14$ GeV/c with a trend towards equality with $R_{AA}^{\pi^0}$ for $p_T\sim 20$ GeV. Should $R_{AA}^{\gamma}$ become equal to $R_{AA}^{\pi^0}$, it would imply that the energy loss in the final state is no longer a significant effect for $p_T\gsim 20$ GeV/c and that the equal suppression of direct-$\gamma$ and $\pi^0$ is due to the initial state structure functions. If this were true, it could mean that going to much higher $p_T$ would not be useful for measurements of parton suppression. In this vein, Kari Eskola gave a nice talk on the latest structure functions in nuclei~\cite{EPS08} at this meeting, but wisely declined to present a prediction for the PHENIX direct-$\gamma$ data, which he said he would show as soon as the preliminary data are published. Clearly, improved measurements of both direct-$\gamma$ and $\pi^0$ in the range $10<p_T<20$ GeV/c are of the utmost importance.     

\subsection{The baryon anomaly}
\begin{figure}[!h]
\begin{center}
\includegraphics[width=0.98\linewidth]{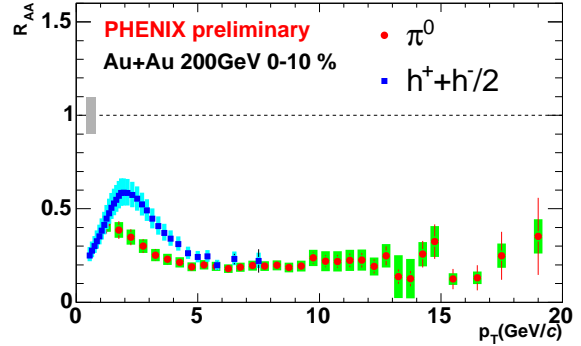} 
\end{center}
\caption[]{$R_{AA}(p_T)$ for $\pi^0$ and non-identified charged hadrons $(h^+ + h^-)/2$ for central (0-5\%) Au+Au collisions at $\sqrt{s_{NN}}=200$ GeV~\cite{MayaQM05}. 
\label{fig:RAApi-h} }
\end{figure}

\begin{figure}[!h]
\begin{center}
\vspace*{+0.03\linewidth}
\includegraphics[width=0.85\linewidth]{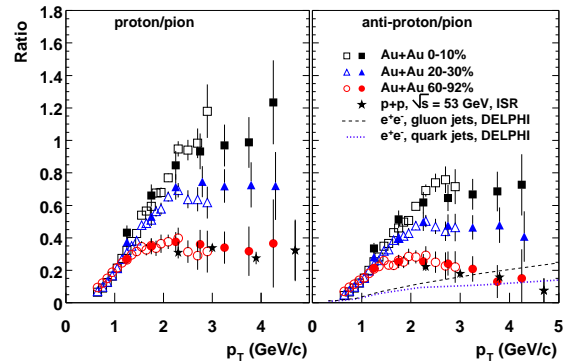} 
\end{center}
\caption[]{$p/\pi^+$ and $\overline{p}/\pi^-$ ratios as a function of $p_T$ and centrality in Au+Au collisions at $\sqrt{s_{NN}}=200$ GeV~\cite{ppg015} compared to other data indicated.
\label{fig:ppi} }
\end{figure}

       There is a tendency of some groups to treat non-identified charged hadrons $h^+$ $h^-$ and correlations among them in A+A collisions as if they were dealing with identified $\pi^0$ mesons. While this might be true in p-p collisions, the situation in Au+Au collisions is quite different as illustrated by Fig.~\ref{fig:RAApi-h}~\cite{MayaQM05}, where $R_{AA}$ for $\pi^0$ and $h^+ + h^-$ are different in the range $2\leq p_T\leq 6$ GeV/c, now called ``intermediate $p_T$.'' Although the effect may appear small on Fig.~\ref{fig:RAApi-h}, when the identified $p/\pi^+$ and $\overline{p}/\pi^-$ ratios were measured in this range (Fig.~\ref{fig:ppi})~\cite{ppg015}, they were an order of magnitude larger than had ever been seen previously in either $e^+ e^-$ jet fragmentation or in the average particle composition of the bulk matter in Au+Au central collisions~\cite{ppg026}.

\begin{figure}[!h]
\begin{center}
\includegraphics[width=0.85\linewidth]{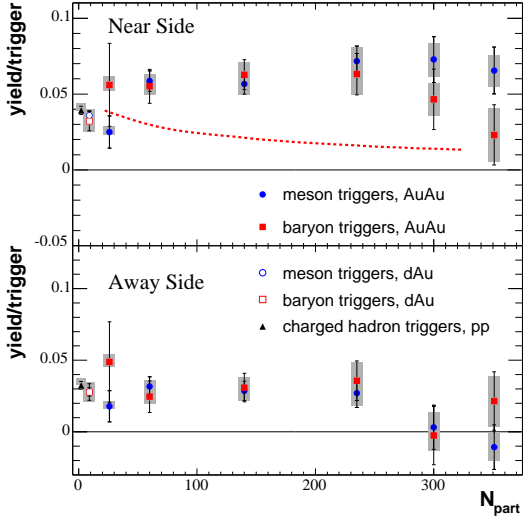} 
\end{center}
\caption[]{Conditional yield per trigger meson (circles), baryon (squares) with $2.5< p_T < 4$ GeV/c, for associated charged hadrons with $1.7 < p_T < 2.5$ GeV/c integrated within $\Delta\phi=\pm 0.94$ radian of the trigger (Near Side) or opposite azimuthal angle, for Au+Au (full), d+Au (open) collisions at $\sqrt{s_{NN}}=200$ GeV~\cite{PXPRC71}. The red-dashed curve indicates the expected trigger-side conditional yield if all the anomalous protons in Au+Au collisions were produced by coalescence. 
\label{fig:Sickles-corr-fig2} }
\end{figure}

   This `baryon anomaly' was beautifully explained as due to the coalescence of an exponential (thermal) distribution of constituent quarks (a.k.a. the QGP)~\cite{GFH03}; but measurements of correlations of $h^{\pm}$ in the range $1.7\leq p_{T_a}\leq 2.5$ GeV/c to identified mesons or baryons with $2.5\leq p_{T_t}\leq 4.0$ GeV/c showed the same near side and away side peaks and yields (Fig.~\ref{fig:Sickles-corr-fig2}) characteristic of di-jet production from hard-scattering~\cite{PXPRC71}, rather than from soft coalescence, apparently ruling out this beautiful model. 

     However, at this meeting, Marco Van Leeuwen showed STAR data presented at QM2008 in which same side correlations with $p_{T_t}>4.0$ GeV/c and $2.0<p_{T_a}<4$ GeV/c are separated into the ridge (large $\delta\eta$ from trigger) and jet region. This result seems to imply that the large $\overline{p}/\pi^-$ ratio $\sim 1$ observed for single inclusive identified particles in the range of the baryon anomaly, $2.0<p_{T}<4.5$ GeV/c, all come from the underlying ridge and not from the smaller jet region! This spectacular observation, which clearly needs to be checked,  opens up a whole host of questions: i) what is the $\overline{p}/\pi^-$ ratio in jets in p-p collisions? (it is $\sim 0.2$ as in jets in $e^+ e^-$ collisions); ii) what are the same and away side correlations in the ridge?; iii) is the ridge the region of equilibrated coalescence? If so why is it localized in azimuth near a jet? iv) why does the azimuthal width of the ridge appear to be similar if not equal to that of a jet? 

\subsection{Heavy quark suppression}
\begin{figure}[!h]
\begin{center}
\includegraphics[width=0.90\linewidth]{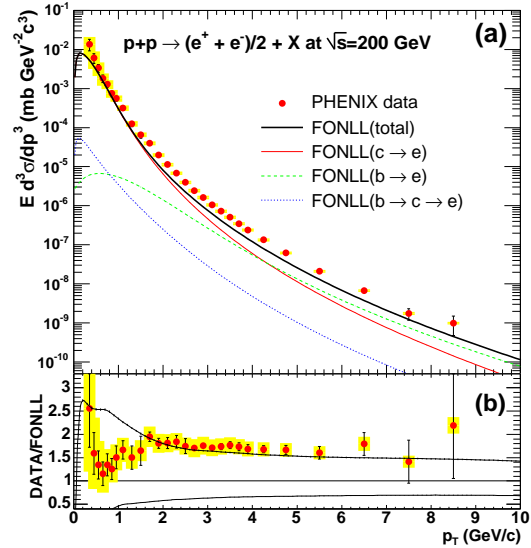} 
\end{center}
\caption[]{a) (top) Invariant cross sections of electrons from heavy flavor decays~\cite{ppg065}. Curves are FONLL theoretical calculations~\cite{ppg065}. b) (bottom) Ratio of the data and the FONLL calculation. The upper (lower) curve shows the theoretical upper (lower) limit of the FONLL calculation.
\label{fig:single-e-pp} }
\end{figure}

\begin{figure}[!h]
\begin{center}
\includegraphics[width=0.90\linewidth]{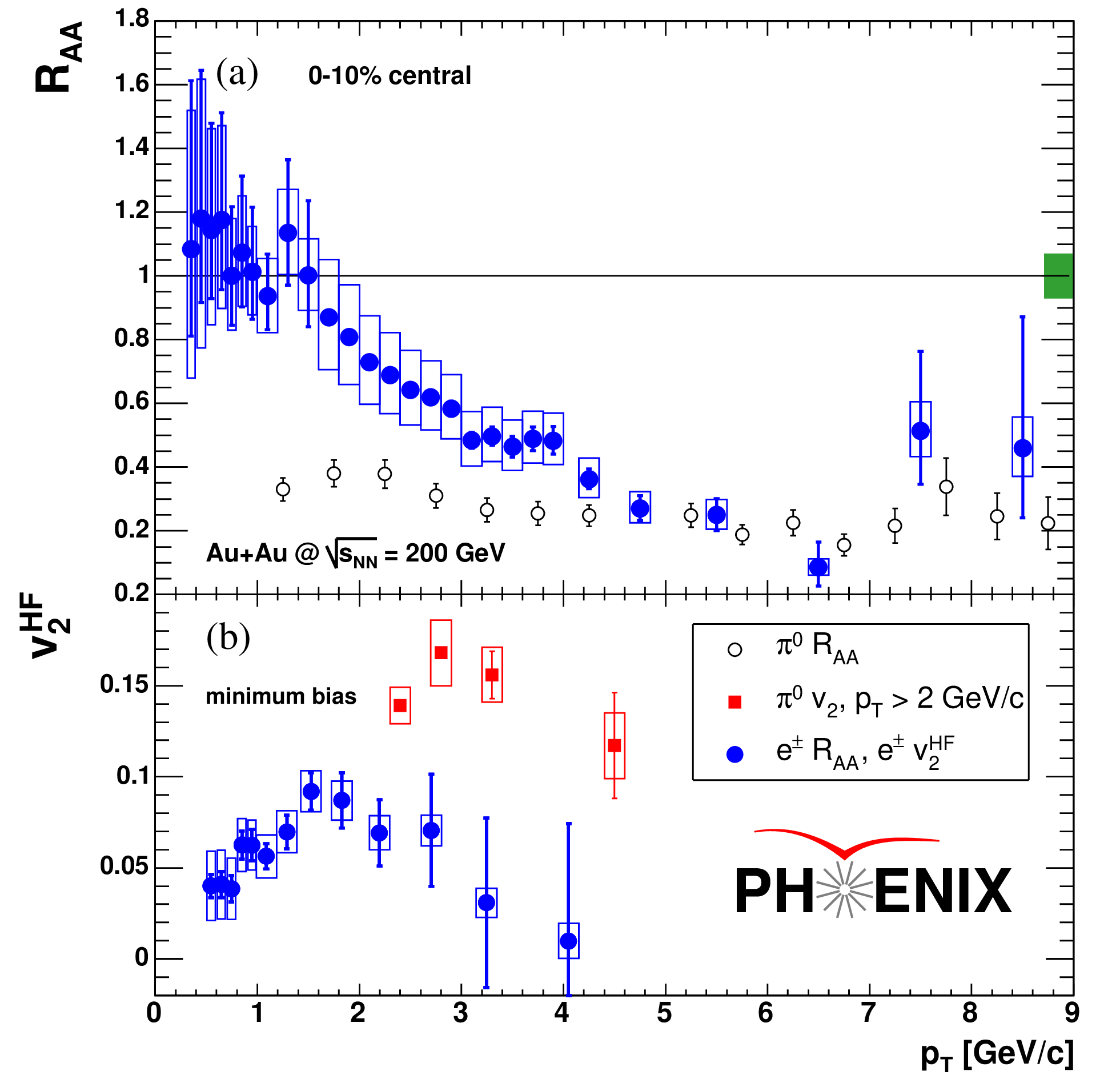} 
\end{center}
\caption[]{a) (top) $R_{AA}$ of heavy-flavor electrons in 0-10\% central collisions compared with $\pi^0$ data. b) Anisotropic flow harmonic $v_2^{HF}$ of heavy-flavor electrons in minimum bias collisions compared with $\pi^0$ data.~\cite{ppg066}.
\label{fig:fig3se} }
\end{figure}

   Another set of striking data at RHIC with no clear explanation at present is the measurement of direct-single $e^{\pm}$ production in p-p collisions (Fig.~\ref{fig:single-e-pp})~\cite{ppg065} in agreement with the FONNL theoretical calculations of semi-leptonic decays of mesons containing $c$ and $b$ quarks,  and the indication by the same measurement in Au+Au collisions of the apparent suppression of heavy quarks $c$ and $b$ by roughly the same amount as $\pi^0$, notably for $p_T\gsim 5$ GeV/c where the $m\gsim 4$ GeV $b$ quarks dominate (Fig.~\ref{fig:fig3se})~\cite{ppg066,seeSTAR}.   

This appears to strongly disfavor the hypothesis of energy loss via gluon bremsstrahlung, which was predicted to be much less for heavy quarks~\cite{deadcone} than for light quarks and gluons; but opens up a whole host of new possibilities including string theory~\cite{egsee066}, as discussed by several talks at this meeting, and even more transformational possibilities~\cite{AZINPC07}. Clearly detailed measurements of correlations of $b-\overline{b}$, $c-\overline{c}$ quarks and light quarks and gluons will be required in order to sort out this very important and very interesting issue.

\subsection{Correlations, jets and fragmentation} 
    The di-jet structure of hard scattering was originally discovered in p-p collisions at the CERN-ISR by measurements of two-particle correlations~\cite{egseeMJTHP04}; and because of the huge multiplicity and the complication of the azimuthal anisotropy due to hydrodynamic flow is the only way that it has been studied so far in A+A collisions at RHIC. 
    
\begin{figure}[!h]
\begin{center}
\includegraphics[width=0.80\linewidth]{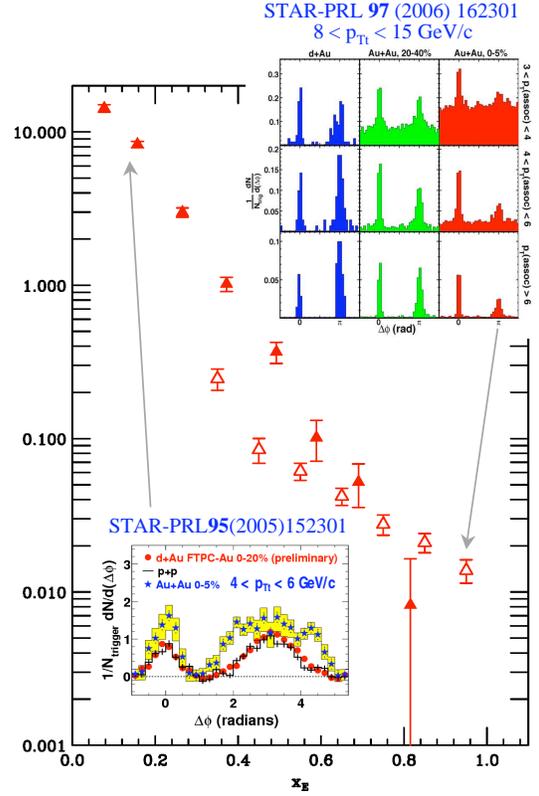} 
\end{center}
\caption[]{Conditional yield of away-side associated $h^{\pm}$ per $h^{\pm}$ trigger with $4< p_{T_t} < 6$ GeV/c (solid points)~\cite{FQWang05} and $8 < p_{T_t} < 15$ GeV/c (open points)~\cite{Magestro} plotted as a function of the ratio of the transverse momentum of the associated particle to the trigger particle $p_{T_a}/p_{T_t}=x_E$. Insets show the conditional probability azimuthal distributions, with flow modulated background subtracted, for both data sets as labeled. 
\label{fig:2stars} }
\end{figure}

    STAR originally claimed that the away-jet vanished in Au+Au collisions~\cite{HardkeQM02} for a trigger $h^{\pm}$ with $4 < p_{T_t} < 6$ GeV/c and associated $h^{\pm}$ with $2< p_{T_a}< p_{T_t}$, but later realized that the away jet didn't vanish it just lost energy and appeared for $h^{\pm}$ with $0.15< p_{T_a}< 4$ GeV/c as a much broader away-side correlation in Au+Au than in p-p collisions~\cite{FQWang05}. The situation was further complicated by the appearance of a narrow away-side peak at still higher $p_{T_t}$ (Fig.~\ref{fig:2stars})~\cite{Magestro}. 
    
\begin{figure}[!h]
\begin{center}
\includegraphics[width=0.90\linewidth]{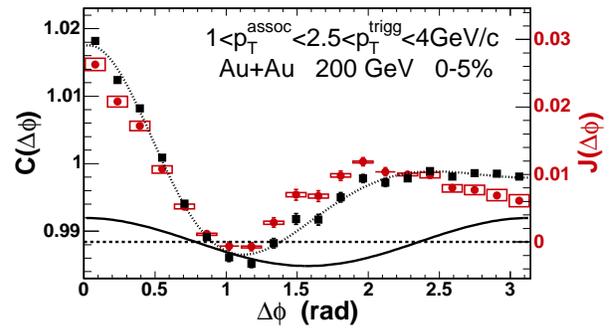} 
\end{center}
\caption[]{Conditional yield azimuthal correlation function, $C(\Delta\phi)$ (black squares), flow background  (solid line) and Jet function $J(\Delta\phi)$ (red dots) after flow subtraction, per trigger $h^{\pm}$ with $2.5< p_{T_t}<4$ GeV/c for associated $h^{\pm}$ of $1.0 < p_{T_a}<2.5$ GeV/c from PHENIX~\cite{ppg067}. PHENIX discusses the half-width $D$ ($\sim 1.1$ radian) of the Jet function $J(\Delta\phi)$ as the angular distance of the apparently displaced peak of the distribution from the angle $\Delta\phi=\pi$. 
\label{fig:JetExtract} }
\end{figure}

    These features were confirmed by PHENIX~\cite{ppg032} with the added fillip of an apparent dip exactly opposite to the trigger particle azimuth, suggesting a two lobed distribution (Fig.~\ref{fig:JetExtract})~\cite{ppg067} or possibly a Mach cone, Cerenkov radiation or other effect resulting from the reaction of the medium to the passage of a fast parton~\cite{seeRefs3267}. 
 \begin{figure}[!h]
\begin{center}
\includegraphics[width=0.90\linewidth]{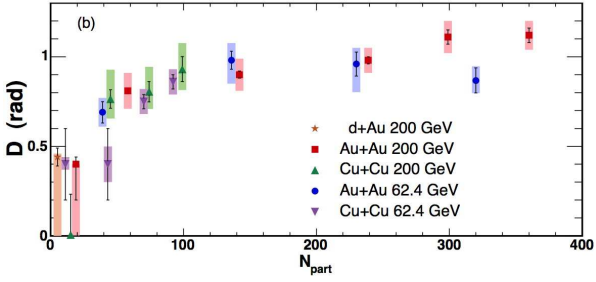} 
\end{center}
\caption[]{PHENIX $D$ parameters~\cite{ppg067} (Fig.~\ref{fig:JetExtract}) as a function of centrality, represented as the number of participants $N_{\rm part}$, for the systems and c.m. energies indicated.
\label{fig:ShapePar-D-only} }
\end{figure}
    One of the striking features of the wide away side correlation is that the width as nicely discussed by Anne Sickles in a talk at this meeting and illustrated by the PHENIX data in Fig.~\ref{fig:ShapePar-D-only} does not depend on centrality, angle to the reaction plane, $p_{T_t}$, $p_{T_a}$ and $\sqrt{s_{NN}}$, which seems problematic to me if the effect is due to a reaction to the medium. Another problematic issue is that all the data upon which the two-lobed correlation function is based are from non-identified $h^{\pm}-h^{\pm}$ correlations in the $p_T$ range of the baryon anomaly where the particle ratios are strongly varying and are anomalous. Another interesting issue seen so far only in a preliminary result from PHENIX (Fig.~\ref{fig:JTMqm06-Azi200Au})~\cite{JTMQM06} is that the same shape away-side correlation persists in Au+Au central collisions even for auto-correlations of particles with very low $p_T$ between 0.2 and 0.4 GeV/c where any effect of hard-scattered partons should be submerged by the predominant soft physics.  Clearly, measurements of correlations with both particles identified and covering a broad range of $p_{T_t}$ and $p_{T_a}$ as a function of the reaction plane are sorely needed. 
\begin{figure}[!h]
\begin{center}
\includegraphics[width=0.90\linewidth]{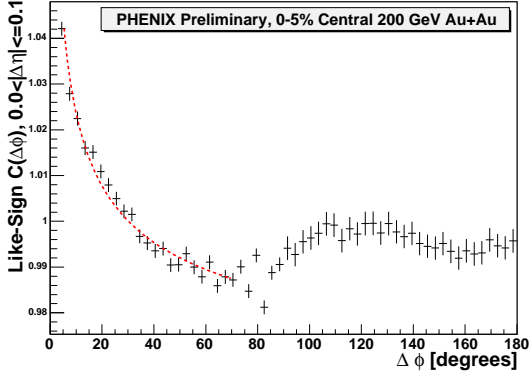} 
\end{center}
\caption[]{Low $p_T$ like-sign pair azimuthal correlation function for 0-5\% central Au+Au collisions at $\sqrt{s}=200$ GeV from charged hadrons with $0.2\leq p_{T_1},\,p_{T_2}\leq 0.4$ GeV/c~\cite{JTMQM06}.
\label{fig:JTMqm06-Azi200Au} }
\end{figure}

\subsubsection{Systematic measurements and punch-through jets}   
          The STAR measurement~\cite{FQWang05} (Fig.~\ref{fig:2stars}) was the first to make a systematic study in $h^{\pm} h^{\pm}$ correlations of the away-side distribution of the ratio of the away-particle to the trigger particle transverse momenta, $p_{T_a}/p_{T_t}$, called $z_T$ by STAR and $x_E$ by PHENIX which was thought to be a determination of the fragmentation function~\cite{FFF}. It was found by PHENIX~\cite{ppg029} that this was not the case, that the away-side $x_E$ distribution triggered by a fragment of a hard-scattered parton was not sensitive to the shape of the fragmentation function of the away-jet but was only sensitive to the power ($n=8.1$) of the semi-inclusive invariant parton $\hat{p}_{T_t}$ spectrum. With no assumptions other than a power law for the parton $\hat{p}_{T_t}$ distribution (${{d\sigma_{q} }/{\hat{p}_{T_t} d\hat{p}_{T_t}}}= A \hat{p}_{T_t}^{-n}$), an exponential fragmentation function ($D^{\pi}_q (z)=B e^{-bz}$), and constant ratio of the away-parton transverse momentum to that of the trigger parton $\hat{x}_h=\hat{p}_{T_a}/\hat{p}_{T_t}$, for fixed $p_{T_t}$ as a function of $p_{T_a}$, it was possible~\cite{ppg029} to derive the $x_E$ distribution in the collinear limit, where $p_{T_a}=x_E p_{T_t}$: 
	     \begin{equation}
\left.{dP_{\pi} \over dx_E}\right|_{p_{T_t}}\approx {N (n-1)}{1\over\hat{x}_h} {1\over
{(1+ {x_E \over{\hat{x}_h}})^{n}}} \, \qquad ,  
\label{eq:condxe2}
\end{equation}
and $N=\mean{m}$ is the multiplicity of the unbiased away-jet. 
                    
          Thus, although not sensitive to the fragmentation function, the $x_E$ ($z_T$) distribution is still sensitive to the ratio of the away parton transverse momentum to the trigger parton transverse momentum, $\hat{x}_h=\hat{p}_{T_a}/\hat{p}_{T_t}$, which is a measure of the differential energy loss of the away parton relative to the trigger parton which is surface biased due to the steeply falling $\hat{p}_{T_t}$ spectrum~\cite{MagestroRef}.  
\begin{figure}[!h]
\begin{center}
\includegraphics[width=0.80\linewidth]{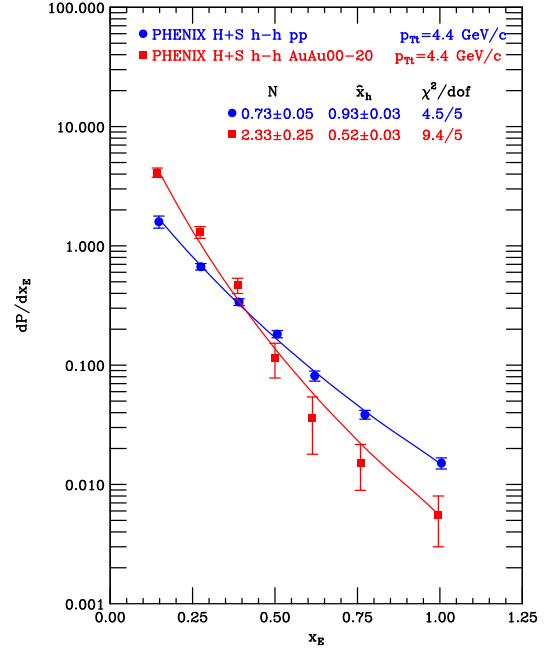} 
\end{center}
\caption[]{Conditional yield of away side ($|\Delta\phi-\pi|<\pi/2$) $h^{\pm}$ per trigger $h^{\pm}$ with $4\leq p_{T_t}\leq 5$ GeV/c~\cite{ppg074} in p-p (circles) and Au+Au central (0-20\%) collisions plotted as $dP/dx_E$ with fits to Eq.~\ref{eq:condxe2} shown and best-fit parameters indicated.
\label{fig:pTt-44-AuAupp} }
\end{figure}
In p-p collisions, the imbalance of the away-parton and the trigger parton  indicated by the fitted value of $\hat{x}_h=0.93\pm 0.03$ in Fig.~\ref{fig:pTt-44-AuAupp} is caused by $k_T$-smearing.  In A+A collisions, the fitted value $\hat{x}_h=0.52\pm 0.03$ indicates that the away parton has lost energy relative to the trigger parton. The fits work well on the PHENIX data so I looked more closely at the two STAR measurements in Fig.~\ref{fig:2stars}. The lower $p_{T_t}$ data set~\cite{FQWang05} nicely followed Eq.~\ref{eq:condxe2} with $\hat{x}_h=0.48$ (see Fig.~\ref{fig:s48cu}) but the higher $p_{T_t}$ data~\cite{Magestro} disagreed in both normalization and shape with the lower $p_{T_t}$ data so I normalized the higher $p_{T_t}$ data to the lower $p_{T_t}$ data in the region $x_E<0.4$ where the slopes seemed to agree and which would be correct if $x_E$ scaling would apply in Au+Au collisions as it does in p-p collisions. When I did this, I was struck by the dramatic break and flattening of the slope in the higher $p_{T_t}$ distribution for $x_E\geq 0.5$. This could be suggestive of a two-component distribution where some partons, which pass through the medium, lose energy, while other partons, such as those emitted tangentially, punch
through without any energy loss. However it is difficult to understand why the punch-through of tangential partons would depend on the trigger $p_{T_t}$. I suggested that the comparison of the two STAR measurements and the possibility of a dramatic break in the $x_E$ distribution would be greatly clarified if a few lower $x_E$ points could be obtained for the higher $p_{T_t}$. STAR presented such a set of preliminary results at Quark Matter 2006 (Fig.~\ref{fig:Horner-AwayPlot-AuAu})~\cite{Horner} which in my opinion show a clear break for $z_T>0.5$ in the range $6<p_{T_t}< 10$ GeV/c, which I believe could represent punch-through of partons which have not lost energy by coherent LPM gluon radiation, but only by standard Bethe-Heitler gluon radiation which presumably is a much smaller effect since it is not coherent. This is a pretty striking observation and a pretty wild guess, so I am surprised and a bit disappointed that there is very little discussion of the `break' in the community. I hope this changes by the next Hard Probes Conference.

\begin{figure}[!h]
\begin{center}
\includegraphics[width=0.80\linewidth]{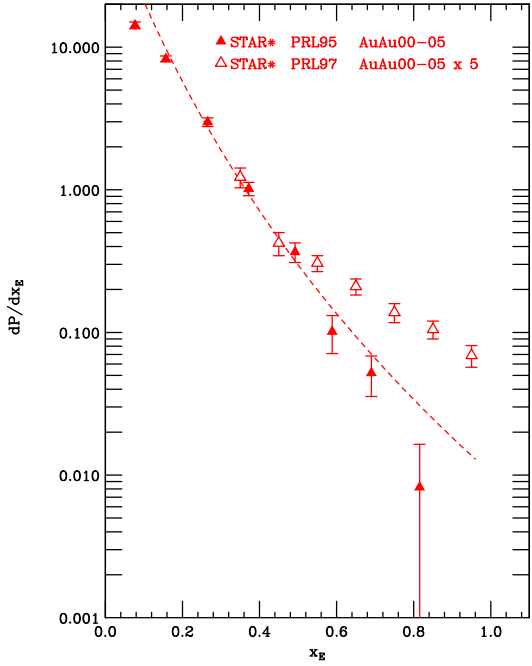} 
\end{center}
\caption[]{$x_E$ distributions from Fig.~\ref{fig:2stars}, with higher $p_{T_t}$ data normalized to agree with lower $p_{T_t}$ data for $x_E<0.4$. Dashed line is a fit of lower $p_{T_t}$ data to Eq.~\ref{eq:condxe2} as described in the text. 
\label{fig:s48cu} }
\end{figure}

 \begin{figure}[!h]
\begin{center}
\includegraphics[width=0.80\linewidth]{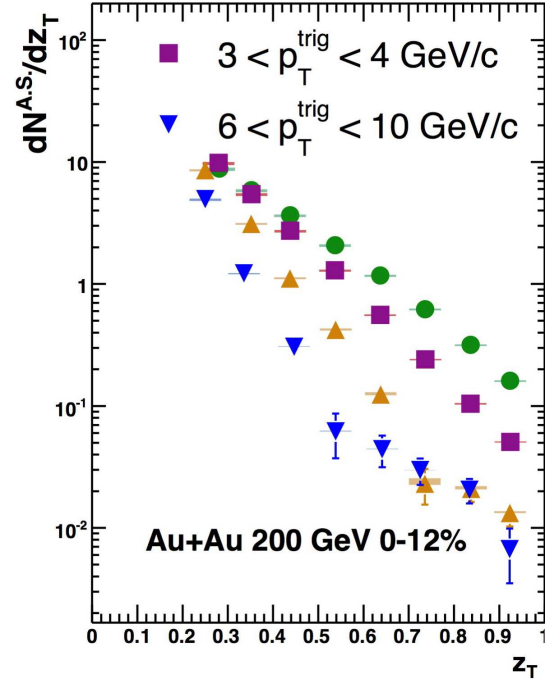} 
\end{center}
\caption[]{STAR $z_T$ ($x_E$) distributions in $h^{\pm}-h^{\pm}$ correlations for 4 intervals of $p_{T_t}$: $2.5<p_{T_t}< 3$ GeV/c (green circles) to $6< p_{T_t}<10$ GeV/c (blue inverted triangles)~\cite{Horner}.
\label{fig:Horner-AwayPlot-AuAu} }
\end{figure}

\subsubsection{Medium modification of jet fragmentation} 
\begin{figure}[!h]
\begin{center}
\includegraphics[width=0.99\linewidth]{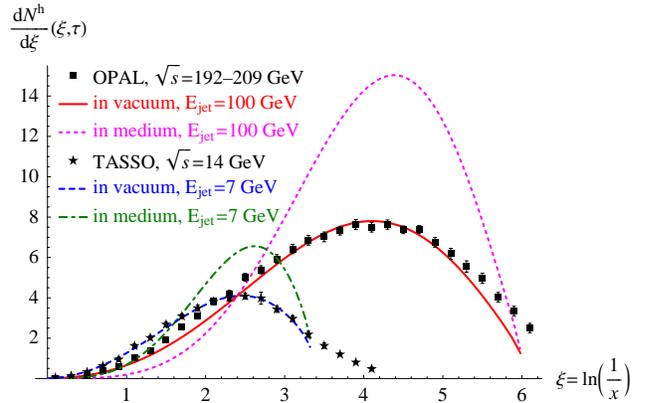} 
\end{center}
\caption[]{Single inclusive distribution of $h^{\pm}$ as a function of $\xi$ for jets measured in $e^+ e^-$ collisions at two values of $\sqrt{s}$ together with MLLA calculations in vacuum and in medium~\cite{BW06}
\label{fig:BorgWied} }
\end{figure}

Borghini and Wiedemann (Fig.~\ref{fig:BorgWied})~\cite{BW06} proposed using the hump-backed or $\xi=\ln(1/z)$ distribution of jet fragments, which is a signature of QCD coherence for small values of  particle momentum fraction, $z=p/E_{\rm jet}$, to explore the medium-modification of jets in heavy ion collisions. The use of the $\xi$ variable would emphasize the increase in the emission of fragments at small $z$ due to the medium induced depletion of the number of fragments at large $z$. The jet energy must be known for this measurement so that it was presumed that full jet reconstruction would be required. 

\begin{figure}[ht] 
\begin{center}
\begin{tabular}{c}
\includegraphics[width=0.75\linewidth]{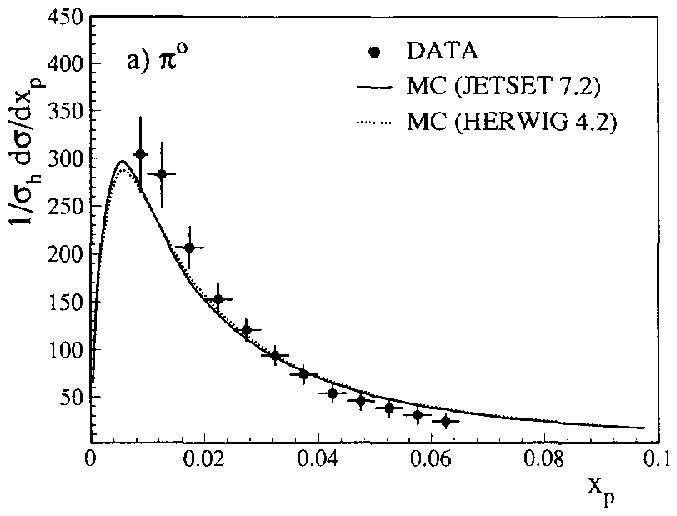}\cr
\includegraphics[width=0.75\linewidth]{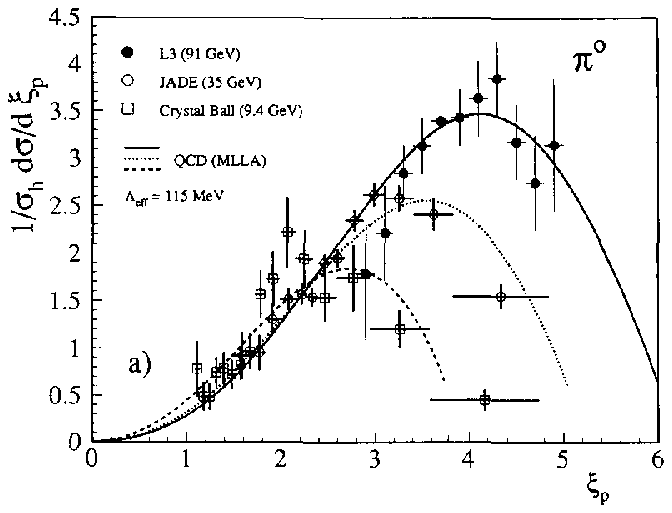}
\end{tabular}
\end{center}
\caption[]
{L3 measurement~\cite{L3} of the inclusive $\pi^0$ spectrum on the $Z^0$ resonance presented as either (top) $x_p=2 p^{\pi^0}/\sqrt{s}$ or (bottom) $\xi_p=\ln (1/x_p)$. 
\label{fig:Ting} }
\end{figure}

	However, one of the original measurements of the $\xi$ distribution in $e^+ e^-$ collisions on the $Z^0$ resonance at LEP was made using the inclusive distribution of $\pi^0$, which could be plotted in either the $z$ or the $\xi$ variable since the energy of the jets for di-jet events was known (Fig.~\ref{fig:Ting})~\cite{L3}. 

A similar state of affairs exists for direct-$\gamma$-hadron correlations in p-p and A+A collisions since, modulo any $k_T$ effect, the jet recoiling from a direct-$\gamma$ has equal and opposite transverse momentum to the precisely measured $\gamma$. Also since the direct-$\gamma$ is a participant in the tree-level partonic reaction $q+g\rightarrow \gamma +q$ and not a fragment, the $x_E$ or $z_T$ distribution of the away-side hadrons from a direct-$\gamma$ actually does represent the away-jet fragmentation function, as suggested by Wang, Huang and Sarcevic~\cite{WHS} so that the $\xi$ distribution can be derived.   

\begin{figure}[ht] 
\begin{center}
\begin{tabular}{c}
\includegraphics[width=0.75\linewidth]{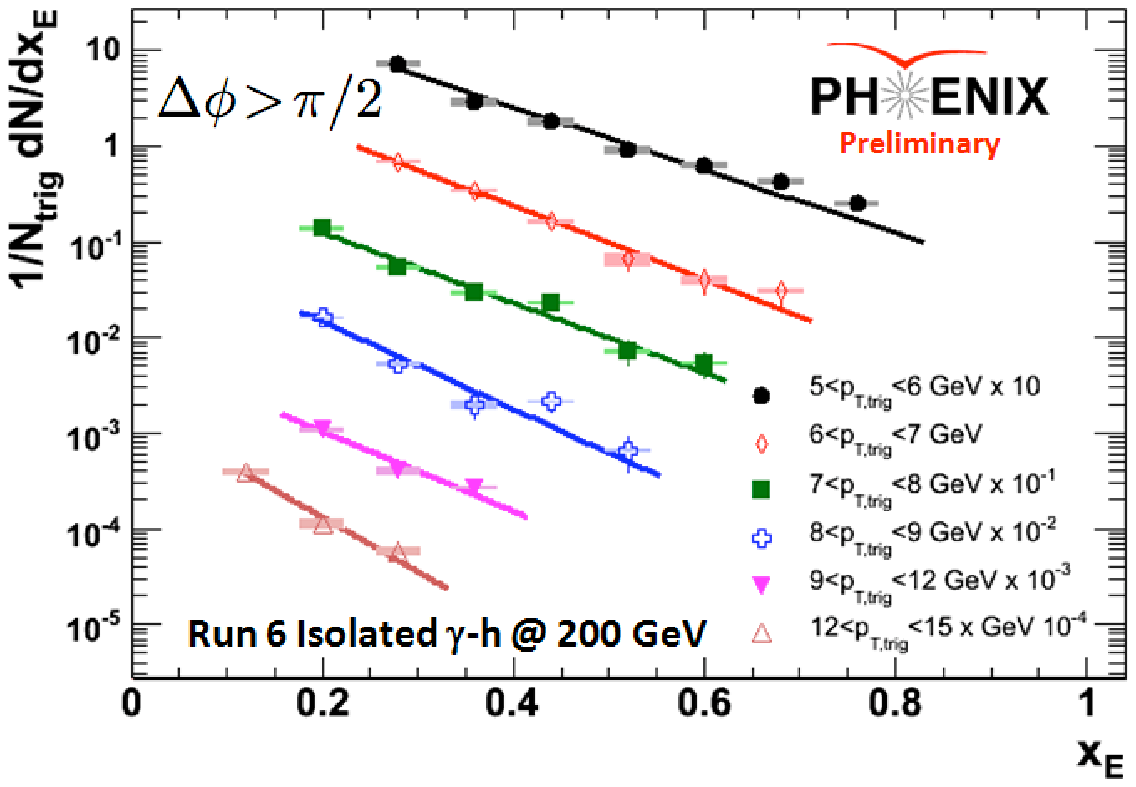}\cr
\includegraphics[width=0.75\linewidth]{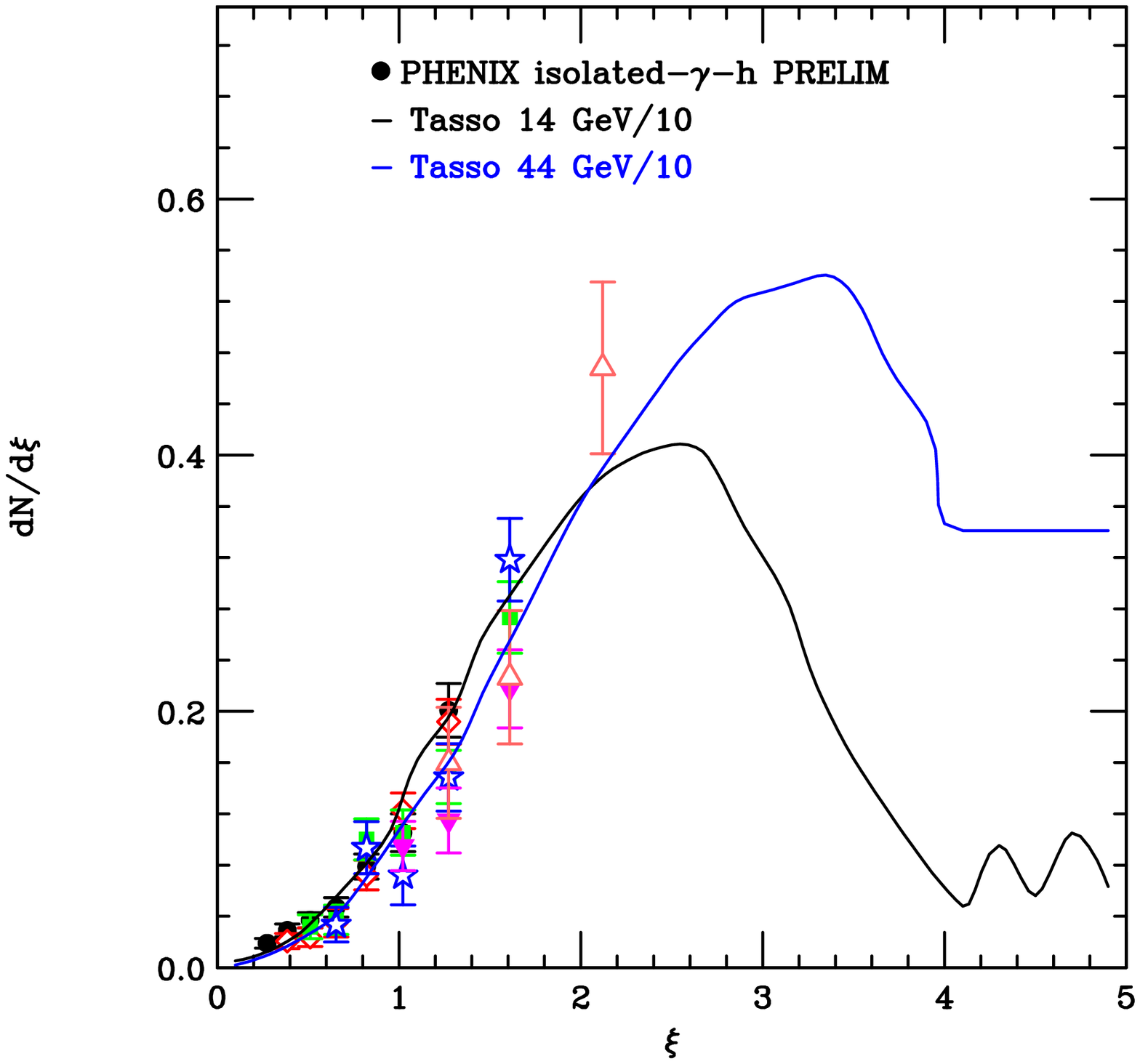}
\end{tabular}
\end{center}
\caption[]
{a) (top) Preliminary PHENIX isolated-direct-$\gamma$ $x_E$ distributions for several ranges of isolated-direct-$\gamma$ $p_{T_t}$ as presented by Justin Frantz at this meeting; b) Same data (using the same symbols) plotted as a function of $\xi=\ln (1/x_E)$ compared to TASSO measurements in $e^+ +e^-$ at two values of $\sqrt{s}$~\cite{TASSO}.   
\label{fig:PXxi} }
\end{figure}

    Justin Frantz showed the preliminary PHENIX isolated-direct-$\gamma$ data from p-p collisions in his talk (see Fig.~\ref{fig:PXxi}a), so I was able to calculate the $\xi$ distribution from this data by the simple change of variables, $dN/d\xi=z\,dN/dz$ (Fig.~\ref{fig:PXxi}b). The PHENIX data nicely follow the trend of the TASSO measurements in $e^+ e^-$ collisions~\cite{TASSO}.    
The $\xi$ distribution clearly emphasizes the fragmentation function in the region $z<0.05$ ($\xi>3.0$) at the expense of the region $z>0.05$, while  
in my opinion it is easier to understand the energy loss of partons in the medium by looking at the fragmentation functions in the standard fragmentation variable $z$, the fractional energy of the jet carried by a fragment particle, as in Fig.~\ref{fig:PXxi}a or the $x_E$ ($z_T$) variable as in Figs.~\ref{fig:2stars}, \ref{fig:pTt-44-AuAupp},~\ref{fig:s48cu},~\ref{fig:Horner-AwayPlot-AuAu}. However, just in case the $\xi$ plot made from direct-$\gamma$-hadron correlations happens to come into common usage, I `modestly' give it the name: Tannenbaum-Ting-Borghini-Wiedemann-Wang plot (in almost alphabetical order).     
\section{``Nobel Dreams'' Redux}
   The appearance of monojets in hard scattering at RHIC, especially in p+A or d+A collisions, has been predicted as a signature of gluon saturation at low $x$~\cite{KLM05}. Due to a famous but erroneous measurement at the SpS collider~\cite{ND} the term and concept of monojets has a very negative connotation to people of my generation.  PHENIX~\cite{PXdAu06} has seen no evidence for mono-jets in $h^\pm -h^\pm$ correlations for triggers with $2.5< p_{T_t} <4$ GeV/c at $\mean{\eta}$=1.7, 0, -1.7, and associated particles with $1.0< p_{T_a} < 2.5$ GeV/c at mid-rapidity ($\mean{\eta}=0$). The widths and conditional yields are the same for triggers at all three values of $\mean{\eta}$ in both p-p and d+Au collisions. On the other hand, STAR~\cite{STARdAu06}, for $\pi^0$ triggers with $1 < \mean{p_{T_t}} <1.4$ GeV/c at $\mean{\eta}=4$ and associated $h^{\pm}$ with $p_{T_a}>0.50$ GeV/c  at $|\eta|<0.75$, appears to see a reduction of both the width and magnitude of the away-side correlation. This situation must be resolved by new data from both p-p and d+Au collisions.  One important issue of concern to me is the background from diffraction dissociation which may be large and even coherent in d+Au collisions in the low $p_{T_t}$ range studied. Also, other, more conventional, pQCD with $k_T$-broadening mechanisms~\cite{QiuVitev} have been proposed which would give a similar effect.  
   
   Clearly, measurements covering a wide range of $\eta_1$, $\eta_2$ and $p_{T_t}$ must be performed in order to verify such an important proposed effect, and I note in this regard that the kinematics for obtaining low $x_2$ are much more favorable with both particles at large $\eta$ since:
   \begin{equation}
   x_1=x_T \frac{e^{\eta_1} + e^{\eta_2}}{2} \qquad x_2=x_T \frac{e^{-\eta_1} + e^{-\eta_2}}{2}  \label{eq:2kin}
   \end{equation} 
   where $x_T=2 p_T/\sqrt{s}$. 

\section{Some of what I think is still not understood} 
To close, I make a list of some of the issues that I think are still not understood at present. With the imminent startup of the LHC, I would not like to bet whether the list gets shorter or longer in the future. 
	\begin{itemize}
\item  Is the nuclear modification factor $R_{AA}$ for $\pi^0$ really constant at a factor of 5 suppression over the range $3< p_T< 20$ GeV/c which would occur for a constant-fractional energy loss analogous to bremsstrahlung, or does the suppression tend to vanish at larger $p_T$? Is dE/dx constant, or a constant fraction, or something else?
\item Does $R_{AA}$ for direct-$\gamma$ really approach that of $\pi^0$ at large $p_T\sim20$ GeV/c as indicated by preliminary data? If true this would argue that the suppression due to a medium effect vanishes at large $p_T> 20$ GeV/c and the effect observed is due to shadowing of the structure functions. If this is confirmed, it would be VERY BAD for LHC.
\item The detailed mechanism of jet suppression due to interaction with the medium is not understood. It is not known whether partons lose energy continuously or discretely; whether they stop in the medium so that the only observed jet fragments are those emitted from the surface; or whether partons merely lose energy exiting the medium such that those  originating from the interior of the medium with initially higher $p_T$ are submerged (due to the steeply falling $p_T$ spectrum) under partons emitted at the surface which have not lost energy. In either case, there is a surface bias. 
\item The reason why heavy quarks appear to lose the same energy as light quarks is not understood.
\item It is not known whether a parton originating at the center of the medium can exit the medium without losing any energy. 
\item It is not known where the energy from the absorbed jets or the parton energy loss goes or how it is distributed. 
\item The surface bias discussed above complicates the use of two-particle correlations of hard-scattered partons to probe the medium since detecting a particle from an away-side parton changes the surface bias of the trigger parton. This means that detection of both a trigger and away side particle is required in order to constrain the hard-scattering kinematics and the position of the origin of the hard-scattered parton-pair within the nuclear matter. Then, the main correlation information with relatively stable kinematics and origin is obtained by studying correlations with an additional 1 or two particles, i.e. a total of 3 or 4 particle correlations, which is much more complicated and requires much more data than the same studies in p-p collisions. 
\item The baryon anomaly, the increase of the p$^{\pm}/\pi^{\pm}$ ratio in the range $2<p_T <6$ GeV/c in Au+Au collisions from the value given by parton-fragmentation in this $p_T$ range in p+p collisions, is not understood. Elegant recombination models fail to explain the similar jet activity  correlated to the p and $\pi$ triggers in this ``intermediate'' $p_T$ range.  
\item The wide away-side non-identified hadron correlations for triggers in the intermediate range $2<p_T <6$ GeV/c in Au+Au collisions, with a possible dip at 180$^o$ which causes apparent peaks displaced by $\sim 60^o$, is not understood. It could represent a Mach cone due to the analogy of a sonic-boom of the parton passing through the medium faster than the speed of sound, or it could indicate jets with large deflections. The effect may be related to the baryon anomaly, which occurs in this $p_T$ range; or the peaks, which are seen also for much softer trigger particles, may not be a hard-scattering effect; or they could represent something totally new.  
\item The ridge is not understood. What causes it? What are its properties? How does it depend on $p_{T_t}$, angle to the reaction plane, centrality, etc? Why isn't there an away-side ridge? How can such a long range correlation $\delta\eta\sim \pm 5$ be created? Is the ridge really the region of the famous equilibrated coalescence with an anomalous $\overline{p}/\pi^-$ ratio? If so why is this region localized near a jet in azimuth and not distributed uniformly in the bulk medium?
\item Are there really mono-jets in d+Au collisions at RHIC energies as predicted by Gluon Saturation?
\item Finally, $J/\Psi$ suppression, which for more than 20 years has represented the gold-plated signature of deconfinement, is not understood.
\end{itemize}

%
%

\end{document}